\documentclass{ifacconf}

\usepackage{graphicx}      
\usepackage{natbib}        
\usepackage{amssymb}          
\usepackage{amsmath}          
\usepackage{enumerate}
\usepackage{algorithm}
\usepackage{algpseudocode}
\usepackage{booktabs}
\usepackage{multirow}
\usepackage{subfigure}

\begin{document}
\begin{frontmatter}

\title{Implementation of real-time moving horizon estimation for robust air data sensor fault diagnosis in the RECONFIGURE benchmark\thanksref{footnoteinfo}} 

\thanks[footnoteinfo]{This work has received funding from the European Union's Seventh Framework Programme (FP7-RECONFIGURE/2007-2013) under grant agreement no. 314544.}

\author[First]{Yiming Wan} 
\author[First]{Tamas Keviczky} 

\address[First]{Delft University of Technology,
	2628CD, Delft, The Netherlands (e-mail: y.wan@tudelft.nl; t.keviczky@tudelft.nl).}

\begin{abstract}                
This paper presents robust fault diagnosis and estimation for the calibrated airspeed and angle-of-attack sensor faults in the RECONFIGURE benchmark. We adopt a low-order longitudinal kinematic model augmented with wind dynamics. In order to enhance sensitivity to faults in the presence of winds, we propose a constrained residual generator by formulating a constrained moving horizon estimation problem and exploiting the bounds of winds. The moving horizon estimation problem requires solving a nonlinear program in real time, which is challenging for flight control computers. This challenge is addressed by adopting an efficient structure-exploiting algorithm within a real-time iteration scheme. Specific approximations and simplifications are performed to enable the implementation of the algorithm using the Airbus graphical symbol library for industrial validation and verification. The simulation tests on the RECONFIGURE benchmark over different flight points and maneuvers show the efficacy of the proposed approach.
\end{abstract}


\end{frontmatter}

\section{Introduction}
In aircraft applications, the industrial state-of-the-art for sensor fault detection and isolation (FDI) relies on triplex hardware redundancy, and performs a majority voting scheme to compute a consolidated value by discarding any failed sources \citep{Goup2015}. This scheme works well if only one sensor source becomes faulty, but it would be inadequate to address simultaneous multiple sensor faults within the triplex redundancy.
As currently investigated in the RECONFIGURE project \citep{Goup2015}, one possibility to extend guidance and control functionalities for future aircraft could be the incorporation of analytical redundancy to detect, isolate, and estimate sensor faults without adding new redundant sensors. 

A crucial issue with the analytical redundancy based FDI technique in aircraft applications is how to simultaneously maintain its robustness to wind disturbances and optimize its fault sensitivity \citep{Marz2012}. 
In \cite{Patton1992}, a robust fault detection approach based on eigenstructure assignment was proposed for faulty sensors of jet engines.
In \cite{Eykeren2014,Hansen2014}, an extended Kalman filter based FDI method was proposed based on the assumption of constant winds. 
Without any limiting assumption about wind dynamics, the disturbance decoupling method based on differential geometry was used in \cite{Cast2014} to perfectly decouple the wind effect in the generated residual signal. 
A similar geometric FDI method was also discussed in \cite{Vanek2011} to deal with multiplicative model uncertainties in an aircraft example.
In \cite{Marcos2005}, a robust $H_\infty$ FDI filter was proposed to estimate the fault signal while attenuating the disturbance effect for a transport aircraft. 
A sliding mode linear parameter varying fault reconstruction scheme was applied in \cite{Alwi2014} to yaw rate sensor faults with extensive industrial validation and verification (V\&V).  A combination of the nullspace based residual generator and the signal-based method were validated in \cite{Daniel2016} for the angle-of-attack (AOA) sensor faults in the RECONFIGURE project.

In our previous work \citep{Wan2016, Wan2016-arxiv}, we have proposed a constrained residual generator to enhance fault sensitivity for simultaneous AOA and calibrated airspeed (VCAS) sensor fault diagnosis in the presence of winds. The constrained residual generator is formulated as a constrained moving horizon estimation (MHE) problem which requires solving a nonlinear program in real time. Compared to conventional unconstrained residual generators, the incorporation of constraints in residual generation further improves fault sensitivity without sacrificing disturbance robustness.
Other relevant aircraft FDI methods related to the above mentioned robustness issue include, but not limited to,  

This paper presents our continued efforts in implementing our MHE based FDI method proposed in the above reference for real-time computation on flight control computers (FCCs). This is a challenging task because computationally intensive nonlinear programming has to be solved within a short sampling interval. Furthermore, for industrial V\&V, the proposed method needs to be implemented using the Airbus graphical symbol library SAO \citep{Briere1993,Fern2015}. This SAO library includes a significantly limited set of mathematical operation blocks, e.g., no single block for matrix manipulations, thus making the implementation of advanced computationally intensive algorithms even more difficult. Our algorithm implementation adopts a real-time iteration scheme with interior-point (IP) sequential quadratic programming (SQP) strategies. It ensures fixed computational cost per sample by limiting the number of iterations and admitting suboptimal solutions. Specific approximations and simplifications are taken to speed up computation.
This implementation has been successfully tested in a number of scenarios using the high-fidelity nonlinear RECONFIGURE benchmark. Its real-time applicability is preliminarily confirmed by Airbus, although future work is needed to further reduce its computational cost. 

\section{Benchmark problem}
One of the objectives we focus on in the RECONFIGURE project is to exploit analytical redundancy to address simultaneous multiple AOA and VCAS sensor faults, which cannot be easily handled by the conventional triplex monitoring technique. The purpose is to detect and isolate any faulty AOA and VCAS sensors, and at the same time, to provide reliable estimation of AOA and VCAS. Table \ref{tab:scenarios} lists eight benchmark scenarios considered in this paper. They include 4 different maneuvers, and cover 8 flight points at different combinations of low or high altitudes and speeds. 
Horizontal and vertical winds are present in all the considered scenarios. 
The fault types include oscillation, jamming, runaway, bias, and non-return to zero (NRZ). Examples of these fault signal profiles can be found in \cite{Goup2015}. In all the listed fault cases, faults occur simultaneously in two of the three redundant sensors of AOA or VCAS. 

There are mainly two constraints when solving the above problem. Firstly, it is considered that during the presence of AOA or VCAS sensor faults, all the other air data sensors probably become unreliable, thus should not be used in the FDI method. 
Secondly, the proposed algorithm should be implemented with the Airbus graphical symbol library SAO for industrial V\&V. This library includes a very limited set of mathematical operation blocks, thus imposing a strict constraint on the complexity level of the implemented algorithm \citep{Briere1993}. Under these constraints, the proposed FDI method should have fast fault detection, very low rate of false alarms and missed detections in the presence of wind disturbances, and low computational cost.


\begin{table*}[ht]
	\caption{Test scenarios. (``Yes'' and ``No'' represents the existence and non-existence of the horizontal/vertical winds. ``F'' and ``H'' stand for ``Faulty'' and ``Healthy'', respectively.)}\label{tab:scenarios}
	\centering
	\begin{tabular}{ccccccccccccc}
		\toprule
		 \multirow{2}{*}{\#} & \multirow{2}{*}{Maneuver}  & Altitude  & Speed  & Horizontal  & Vertical  & \multirow{2}{*}{Fault type} & \multicolumn{3}{c}{AOA sensors} & \multicolumn{3}{c}{VCAS sensors} \\
		 &           & [ft]      & [kts] & wind & wind &  & $\alpha_m^{(1)}$ & $\alpha_m^{(2)}$ & $\alpha_m^{(3)}$ & $V_{c,m}^{(1)}$ & $V_{c,m}^{(2)}$ & $V_{c,m}^{(3)}$\\
		\midrule
		1 & \multirow{2}{*}{Load factor control} & 7475 & 207 & Yes & No & NRZ & F & F & H & F & F & H \\
		2 & & 1565 & 203 & Yes & No & Oscillation & F & F & H & F & F & F \\
		3 & \multirow{2}{*}{Flight path angle control} & 15896 & 277 & Yes & Yes & Oscillation & F & F & H & F & F & H \\
		4 & & 31331 & 322 & Yes & Yes & Bias & H & H & H & F & F & H \\
		5 & \multirow{2}{*}{Vertical speed control} & 30666 & 332 & Yes & Yes & Jamming & H & H & H & F & F & H \\
		6 & & 15658 & 351 & Yes & Yes & Runaway & H & H & H & F & F & H \\
		7 & \multirow{2}{*}{AOA protection} & 943 & 162 & Yes & Yes & Jamming & F & F & H & F & F & H \\
		8 & & 968 & 173 & Yes & Yes & NRZ & H & H & H & F & F & H \\
		\bottomrule
	\end{tabular}
\end{table*}

\section{Modeling of Longitudinal Motions and wind dynamics}\label{sect:mdl}
Since only longitudinal dynamics is investigated in this paper, the following model is adopted:
\begin{equation}\label{eq:ct_ss}
	\left\{
	\begin{aligned}
		{\mathbf{\mathbf{\dot \alpha}}} (t) &= {f} \left( \alpha (t), \varTheta (t) \right) + {u_{\alpha}}(t)  \\
		\mathbf{\dot w}(t) &= \mathbf{u}_w (t)  \\
		{\mathbf{y}} (t) &= h \left( \alpha (t), \mathbf{w} (t), \varTheta (t) \right) \\
		{\mathbf{y}}_m (t) &= {\mathbf{y}} (t) + \mathbf{n}(t)
	\end{aligned}
	\right.
\end{equation}
with the definitions $\varTheta = \left[ \begin{matrix}
V_g & \theta & q & n_x & n_z & z
\end{matrix} \right]^\mathrm{T}$, 
$\mathbf{w} = \left[ \begin{matrix}
W_x & W_z
\end{matrix} \right]^\mathrm{T}$,
$\mathbf{u}_w = \left[ \begin{matrix}
u_{w,x} & u_{w,z}
\end{matrix} \right]^{\mathrm{T}}$,
$\mathbf{y} = \left[ \begin{matrix}
\alpha & V_z & V_{c}
\end{matrix} \right]^\mathrm{T}$, and
$\mathbf{n} = \left[ \begin{matrix}
n_{\alpha} & n_{vz} & n_{vc}
\end{matrix} \right]^\text{T}$.
The system outputs ${\mathbf{y}} (t)$ include AOA $\alpha$, vertical speed $V_z$, and VCAS $V_c$. 
The output measurements $\mathbf{y}_m (t)$ are corrupted with white Gaussian measurement noises $\mathbf{n}(t)$.
$W_x$ and $W_z$ represent horizontal and vertical wind speeds, respectively. 
The model parameter $\varTheta$ consists of ground speed $V_g$, pitch angle $\theta$, pitch rate $q$, horizontal load factor $n_x$, vertical load factor $n_z$, and altitude $z$, which are all measurable.
The input noise $u_\alpha$ accounts for the model mismatch.
The output equations in (\ref{eq:ct_ss}) for $V_z$ and 
$V_c$ are $V_z = h_{vz} (\alpha, \mathbf{w}, \varTheta)$ and
$V_{c} = h_{vc} (\alpha, \mathbf{w}, \varTheta)$,
respectively. For each redundant AOA sensor measurement $\alpha_{m}^{(i)}$ or VCAS sensor measurement $V_{c,m}^{(i)}$, $i = 1,2,3$, the latent sensor faults $f_{\alpha}^{(i)}$ and $f_{vc}^{(i)}$ are additive, i.e.,
\begin{equation*}\label{eq:hvc_f}
	\alpha_{m}^{(i)} = \alpha + f_{\alpha}^{(i)} + n_{\alpha}^{(i)},\;
	V_{c,m}^{(i)} = V_c + f_{vc}^{(i)} + n_{vc}^{(i)}.
\end{equation*}
The first-order integrating model in (\ref{eq:ct_ss}) is a simple yet powerful approximation of the wind dynamics that has been widely used in flight control, e.g., in \cite{Mulg1996}. $u_{w,x}$ and $u_{w,z}$ represent horizontal and vertical wind accelerations.

The system model (\ref{eq:ct_ss}) is adopted due to several considerations: a) it avoids using other air data measurements which are considered as unreliable in the presence of AOA or VCAS sensor faults, and involves only inertial sensors; b) it includes no aerodynamic parameters, avoiding the issue of robustness to uncertain aerodynamic parameters; c) its low state dimensions are attractive for real-time computation. The reader is referred to \cite{Wan2016-arxiv} for more details about the model (\ref{eq:ct_ss}).


\section{Fault-tolerant moving horizon estimation scheme}\label{sect:scheme}
\begin{figure}[!h]
	\centering
	\includegraphics[width=0.55\linewidth]{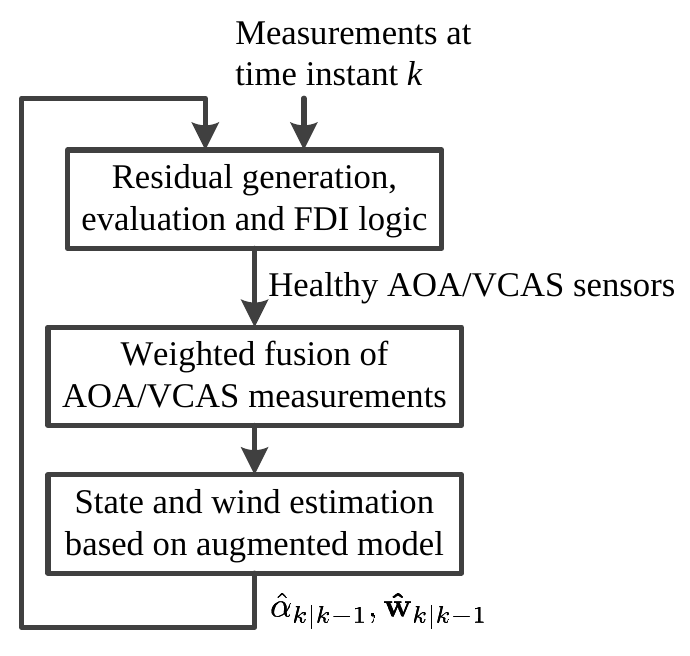}
	\caption{Fault detection and isolation scheme}
	\label{fig:FTE_scheme}
\end{figure}

As depicted in Figure \ref{fig:FTE_scheme}, our FDI scheme consists of three consecutive steps: a) isolating faulty AOA/VCAS sensors based on generated residual signals; b) fusing the healthy AOA and VCAS sensors into two weighted mean values; c) estimating states and winds.
The residual signals for FDI are generated as the difference between the AOA/VCAS measurements $\{ \alpha_{m,k}^{(i)}, V_{c,m,k}^{(i)} \}$ and their one-step-ahead predictions $\{ \hat \alpha_{k|k-1}, \hat V_{c,k|k-1} \}$, i.e.,
\begin{equation*}\label{eq:respred_vc}
\begin{aligned}
r_{\alpha,k}^{(i)} = \alpha_{m,k}^{(i)} - \hat \alpha_{k|k-1}, \,
r_{vc, k}^{(i)} = V_{c,m,k}^{(i)} - \hat V_{c,k|k-1}, \, i=1,2,3.
\end{aligned}
\end{equation*}
Here, the index $k$ denotes the samples at time instant $t_k$. 
The residual signals are evaluated by their root mean square (RMS) values over a sliding window:
\begin{equation}\label{eq:rms}
J_{\star,k}^{(i)} = \sqrt{ \frac{1}{N_{\text{eval}}} \sum_{j=k-N_{\text{eval}}+1}^{k} (r_{\star,j}^{(i)})^2 }
\end{equation}
where $\star$ represents ``$\alpha$'' and ``$vc$'', $N_{\text{eval}}$ is the length of residual evaluation window. With suitable threshold $J_{\star, \mathrm{th}}$, the $i$th AOA or VCAS sensor is concluded to be faulty if we have $J_{\star,k}^{(i)} > J_{\star, \mathrm{th}}$ for $n_d$ times during the past time window $[k - N_{\text{eval}} + 1, k]$, which allows a confirmation time for the detected fault.

The redundant AOA sensors identified as fault-free are fused into a weighted mean measurement $\alpha_m$:
\begin{equation}\label{eq:fuse_w}
\begin{aligned}
\alpha_{m,k} &= \sum_{i \in \{ J_{\alpha, k}^{(i)} \leq J_{\alpha, \text{th}} \}} \beta_{\alpha, k}^{(i)} \alpha_{m,k}^{(i)}, \\
\beta_{\alpha,k}^{(i)} &= \frac{1}{  \sum\limits_{j \in \{ J_{\alpha, k}^{(j)} \leq J_{\alpha, \text{th}} \} } \frac{1}{ \left(J_{\alpha, k}^{(j)} \right)^2} } 
\frac{1}{\left(J_{\alpha, k}^{(i)} \right)^2}.
\end{aligned}
\end{equation}
Similar to \cite{Eykeren2014}, the above adaptive weights $\beta_{\alpha,k}^{(i)}$ are computed from the residual RMS values, so that the sensors with larger residual RMS values are assigned with lower weights.
We follow the same procedure to compute the weights $\beta_{vc,k}^{(i)}$ and the weighted mean value $V_{c,m,k}$ from the VCAS sensors identified as fault-free.
These fused measurements are used in state and wind estimation, as in Figure \ref{fig:FTE_scheme}.
Before a faulty sensor is detected, the undetected faulty sensor is given a lower weight in (\ref{eq:fuse_w}) due to its larger residual RMS value, therefore the weighted fusion reduces the estimation errors caused by fault detection delays.

The state and wind estimation algorithm computes the one-step ahead predictions $\hat \alpha_{k|k-1}$ and $\hat V_{c,k|k-1}$ from the filtered estimates $\hat \alpha_{k-1|k-1}$ and $\mathbf{\hat w}_{k-1|k-1}$ based on the model (\ref{eq:ct_ss}). A nonlinear constrained MHE problem is formulated to exploit the bounds of $\alpha$, $\mathbf{w}$, $u_\alpha$, and $\mathbf{u}_w$ in residual generation, which will be explained in Section \ref{sect:MHE}. It is well known that the incorporation of additional signal information as state constraints in a state estimation problem can improve the estimation accuracy \citep{Simon2010}. Moreover, from the FDI point of view, compared to conventional unconstrained residual generation, the incorporation of constraints in residual generation can improve fault sensitivity without reducing disturbance robustness. The reader is referred to \cite{Wan2016, Wan2016-arxiv} for more detailed analysis.


\section{Moving horizon estimation and its real-time implementation}\label{sect:MHE}
The nonlinear constrained MHE method is the core of our proposed FDI scheme given in Section \ref{sect:scheme}. In this section, we will first fromulate the MHE problem, and then show how it is implemented with SAO for real-time computation.

The MHE framework builds on the discrete-time approximation of the continuous-time model (\ref{eq:ct_ss}):
\begin{subequations}\label{eq:dt_ss}
	\begin{align}
	\alpha_{k+1} &= \alpha_k + t_s f(\alpha_k, \varTheta_k) + t_s {u}_{\alpha,k},    \label{eq:dt_longdyn} \\
	\mathbf{w}_{k+1} &= \mathbf{w}_{k} + t_s \mathbf{u}_{w,k}, \label{eq:dt_winddyn} \\
	\mathbf{\bar y}_{m,k} &= h(\alpha_k, \mathbf{w}_{k}, \varTheta_k) + \mathbf{\bar n}_{k},   \label{eq:dt_longoutput}
	\end{align}
\end{subequations}
where $t_s$ is the sampling interval, (\ref{eq:dt_longdyn}) 
and (\ref{eq:dt_winddyn}) are obtained via approximated numerical integration applied to (\ref{eq:ct_ss}). In (\ref{eq:dt_longoutput}), the output vector $\mathbf{\bar y}_{m,k}$ consists of the vertical speed measurement $V_{z,m}$ and the two fused  measurements $\alpha_m$ and $V_{c,m}$ defined in (\ref{eq:fuse_w}), i.e.,
\begin{equation}\label{eq:ymeas}
\mathbf{\bar y}_{m,k} = \left[ \begin{matrix}
\alpha_{m,k}  & V_{z,m,k} 
& V_{c,m,k} \end{matrix} \right]^\text{T}.
\end{equation}
Then $\mathbf{\bar n}_k$ is defined similarly to (\ref{eq:fuse_w}).

Given a moving horizon including $N$ samples of output measurements $\{ \mathbf{\bar y}_{m,l}, \mathbf{\bar y}_{m,l+1}, \dots, \mathbf{\bar y}_{m,k} \}$ ($l = k-N+1$) at time instant $k$, the MHE problem is formulated as 
\begin{subequations}\label{eq:MHE}
	\begin{align}
	\mathop {\min }\limits_{\begin{smallmatrix}
		\mathbf{x}_{i}, \mathbf{u}_{i} 
		\end{smallmatrix}} \;
	& \frac{1}{2}\left\| {{\mathbf{x}_l} - {\mathbf{\hat x}_{l|k-1}} } \right\|_{\mathbf{P}^{ - 1}}^2
	+ \frac{1}{2}\sum\limits_{i = l}^{k - 1} {\left\| \mathbf{u}_{i} \right\|_{\mathbf{Q}^{-1}}^2}   \label{eq:MHEobj} \\
	& + \frac{1}{2}\sum\limits_{i = l}^k {\left\| {\mathbf{\bar y}_{m,i} - h ({\mathbf{x}_i}, \varTheta_i) } \right\|_{\mathbf{R}_i^{ - 1}}^2}  \nonumber \\
	\text{s.t.} \; & \mathbf{x}_{i+1} = F(\mathbf{x}_i, \mathbf{u}_i, \varTheta_i),  \label{eq:MHEcon} \\
	& \mathbf{u}^{\text{LB}} \leq \mathbf{u}_{i} \leq \mathbf{u}^{\text{UB}}, \; i=l,\dots,k-1, \nonumber \\
	& \mathbf{x}^{\text{LB}} \leq \mathbf{x}_{i} \leq \mathbf{x}^{\text{UB}}, \; i = l,\dots,k, \nonumber
	\end{align}
\end{subequations}
where
\begin{equation}\label{eq:MHEdef}
\begin{aligned}
& \mathbf{x} = \left[ \begin{matrix}
\alpha \\
\mathbf{w}
\end{matrix} \right], 
\mathbf{u} = \left[ \begin{matrix}
u_\alpha \\
\mathbf{u}_w
\end{matrix} \right], 
\mathbf{P} = \left[ \begin{matrix}
p_\alpha & \mathbf{0}\\
\mathbf{0} & p_w \mathbf{I}_2
\end{matrix} \right], \\
&\mathbf{Q} = \left[ \begin{matrix}
q_\alpha & \mathbf{0}\\
\mathbf{0} & q_w \mathbf{I}_2
\end{matrix} \right], 
\mathbf{R}_i = \text{diag}(R_{\alpha,i}, R_{vz}, R_{vc,i}), \\
& R_{\star, i} = \sum_{j \in \{ J_{\star,k}^{(j)} \leq J_{\star, \text{th}} \}  } \left(\beta_{\star,k}^{(j)}\right)^2 R_\star, 
\end{aligned} 
\end{equation}
with $\star$ representing ``$\alpha$'' and ``$vc$''. 
The function $F(\cdot)$ in (\ref{eq:MHEcon}) represents the right-hand side of (\ref{eq:dt_longdyn}) and (\ref{eq:dt_winddyn}).
Detailed explanation about the above MHE formulation is referred to \cite{Rawl2006}. The positive scalars $p_\alpha$, $p_w$, $R_\alpha$, $R_{vz}$, and $R_{vc}$ are tuning parameters in the weighting matrices, and 
${R}_{\star,i}$ takes its form in (\ref{eq:MHEdef}) according to the definition of the fused estimates in (\ref{eq:fuse_w}). These parameters can be tuned for the trade-offs between sensitivity to faults and robustness to disturbances, which has been analyzed in \cite{Wan2016, Wan2016-arxiv}.

To solve the nonlinear programming problem (\ref{eq:MHE}), we adopt a generalized Gauss-Newton (GGN) SQP strategy and use the IP method to solve each quadratic programming (QP) subproblem. For real-time computation within a short sampling interval, we perform only one SQP iteration per sample and fix the number of iterations in solving each QP subproblem. 
This real-time iteration strategy, which has been reported in the literature, e.g., in \cite{Kuhl2011}, admits
sub-optimality of the solution to enable fixed computational cost per sample. 
In the following subsections, we will present the details of its SAO implementation.


\subsection{Generalized Gauss-Newton SQP}
\begin{algorithm}[!h]
	\caption{Primal barrier interior-point algorithm}
	\label{alg:primal_barrier}
	\begin{algorithmic}
		\State \textit{Initialization}: $\kappa = \kappa_{\text{init}}$,
		$\{\Delta \mathbf{x}_{i}^{-}=0, \Delta \mathbf{u}_i^{-} = 0 \}_{i = l, \cdots, k}$.
		\For{$j = 0 \rightarrow n_\kappa n_{QP}-1$}
		\State \textit{Linearization}: compute all the quantities in (\ref{eq:linearize}).
		\State \textit{Compute the search direction} $\{\Delta^2 \mathbf{x}_{i}, \Delta^2 \mathbf{u}_{i} \}_{i=l,\cdots,k}$ with Algorithm \ref{alg:linear_solver}.
		\State \textit{Line search}:
		$$ s_j = \max \left\{ 2^{-n_s} \left| \begin{array}{l}
		\Delta \mathbf{\star}_i^{\text{LB}} \leq \eta_{\star,i}(2^{-n_s}) \leq \Delta \mathbf{\star}_i^{\text{UB}}, \\
		i = l, \cdots, k, \text{ and }n_s \leq n_s^{\max} \\ 
		\text{ is a nonnegative integer}
		\end{array} 
		\right. \right\}
		$$
		where $\eta_{\star,i} (\mu) = \Delta \mathbf{\star}_i^- + \mu \Delta^2 \mathbf{\star}_i$ with $\star$ representing $\mathbf{u}$ and $\mathbf{x}$. $s_j$ is rounded to zero if $\eta_{\star,i} (2^{-n_s^{\max}})$ cannot satisfy all the inequality constraints.
		\State \textit{Update}: \\
		$\quad\; (\Delta \mathbf{x}_i^{-}, \Delta \mathbf{u}_i^{-})  \leftarrow (\Delta \mathbf{x}_i^{-}, \Delta \mathbf{u}_i^{-}) + s_j (\Delta^2 \mathbf{x}_i, \Delta^2 \mathbf{u}_i)$
		\State $\kappa \leftarrow 0.1 \kappa$ if $\text{mod}(j, n_\kappa) = 0$
		\EndFor
		\State \textit{Solution} to the QP (\ref{eq:QP}): $\left(\Delta \mathbf{x}_{i|k}, \Delta \mathbf{u}_{i|k} \right) \leftarrow \left( \Delta \mathbf{x}_{i}^-, \Delta \mathbf{u}_{i}^- \right) $
	\end{algorithmic}
\end{algorithm}

At each time instant $k$, the original problem (\ref{eq:MHE}) is first linearized by applying the GGN SQP strategy, around the solution $\mathbf{\hat x}_{i|k-1}$ ($i=l,\cdots,k$) and $\mathbf{\hat u}_{i|k-1}$ ($i=l,\cdots,k-1$) of the previous time instant $k-1$. Note that $\mathbf{\hat x}_{k|k-1} = F(\mathbf{\hat x}_{k-1|k-1}, \mathbf{\hat u}_{k-1|k-1}, \varTheta_{k-1})$ is the predicted estimate at time instant $k-1$. This leads to the following QP subproblem:
\begin{subequations}\label{eq:QP}
	\begin{align}
	\mathop {\min }\limits_{\begin{smallmatrix}
		\Delta \mathbf{x}_i, \Delta \mathbf{u}_i
		\end{smallmatrix}} \;
	& \frac{1}{2}\left\| {{\Delta \mathbf{x}_l} } \right\|_{\mathbf{P}^{ - 1}}^2
	+ \frac{1}{2}\sum\limits_{i = l}^{k - 1} {\left\| \mathbf{r}_{u,i} - \Delta \mathbf{u}_i \right\|_{\mathbf{Q}^{-1}}^2}   \label{eq:QPobj} \\
	& + \frac{1}{2}\sum\limits_{i = l}^k {\left\| {\mathbf{ r}_{y,i} - \mathbf{C}_i \Delta \mathbf{x}_i } \right\|_{\mathbf{R}_i^{ - 1}}^2}  \nonumber \\
	\text{s.t.} \; & \Delta \mathbf{x}_{i+1} = \mathbf{f}_i + \mathbf{A}_i \Delta \mathbf{x}_{i} + \mathbf{B}_i \Delta \mathbf{u}_{i},  \label{eq:QPcon} \\
	& \Delta \mathbf{u}_i^{\text{LB}} \leq \Delta \mathbf{u}_{i} \leq \Delta \mathbf{u}_i^{\text{UB}}, \; i=l,\dots,k-1, \nonumber \\
	& \Delta \mathbf{x}_i^{\text{LB}} \leq \Delta \mathbf{x}_{i} \leq \Delta \mathbf{x}_i^{\text{UB}}, \; i = l,\dots,k, \nonumber
	\end{align}
\end{subequations}
where
$$\Delta \mathbf{x}_{i} = {\mathbf{x}_i} - {\mathbf{\hat x}_{i|k-1}},\, \Delta \mathbf{u}_{i} = {\mathbf{u}_i} - {\mathbf{\hat u}_{i|k-1}},$$ 
$$\mathbf{r}_{u,i} = - {\mathbf{\hat u}_{i|k-1}},\, \mathbf{r}_{y,i} = \mathbf{\bar y}_{m,i} - h(\mathbf{\hat x}_{i|k-1}, \varTheta_{i}),$$ 
\begin{equation}\label{eq:Ai}
\mathbf{A}_i = \text{diag}( 1 + t_s \nabla_{\mathbf{x}_i} f,1,1),\,
\mathbf{B}_i = t_s \mathbf{I}_{n_u},\,
\mathbf{C}_i = \nabla_{\mathbf{x}_i} h, 
\end{equation}
$$\mathbf{f}_i = \mathbf{\hat x}_{i|k-1} - \mathbf{\hat x}_{i+1|k-1} + t_s F \left( \mathbf{\hat x}_{i|k-1}, \mathbf{\hat u}_{i|k-1} \right), $$
$$ \Delta \mathbf{u}_i^\text{LB} = \mathbf{u}^\text{LB} - \mathbf{\hat u}_{i|k-1}, \,
\Delta \mathbf{u}_i^\text{UB} = \mathbf{u}^\text{UB} - \mathbf{\hat u}_{i|k-1},$$
$$ \Delta \mathbf{x}_i^\text{LB} = \mathbf{x}^\text{LB} - \mathbf{\hat x}_{i|k-1}, \,
\Delta \mathbf{x}_i^\text{UB} = \mathbf{x}^\text{UB} - \mathbf{\hat x}_{i|k-1}.
$$
Its solution $\{\Delta \mathbf{x}_{i|k}, \Delta \mathbf{u}_{i|k} \}$ is computed by using the algorithm given in Section {\ref{sect:QPsub}}. Finally, the solution to the original problem (\ref{eq:MHE}) is updated as
$\mathbf{\hat x}_{i|k} = \mathbf{\hat x}_{i|k-1} + \Delta \mathbf{x}_{i|k}$ and $\mathbf{\hat u}_{i|k} = \mathbf{\hat u}_{i|k-1} + \Delta \mathbf{u}_{i|k}$, and used to initialize the SQP iteration at the next time instant.

\subsection{Solving the QP subproblem}\label{sect:QPsub}
An infeasible start primal barrier IP method is adopted to solve the QP subproblem (\ref{eq:QP}). We first replace the inequality constraints in the QP (\ref{eq:QP}) with barrier terms in its objective function, to get the approximate problem
\begin{align}
	\mathop {\min }\limits_{\begin{smallmatrix}
		\Delta \mathbf{x}_i, \Delta \mathbf{u}_i
		\end{smallmatrix}} \;
	& \frac{1}{2}\left\| {{\Delta \mathbf{x}_l} } \right\|_{\mathbf{P}^{ - 1}}^2
	+ \frac{1}{2}\sum\limits_{i = l}^{k - 1} {\left\| \mathbf{r}_{u,i} - \Delta \mathbf{u}_i \right\|_{\mathbf{Q}^{-1}}^2}  \label{eq:QP_barrier} \\
	&+ \frac{1}{2}\sum\limits_{i = l}^k {\left\| {\mathbf{ r}_{y,i} - \mathbf{C}_i \Delta \mathbf{x}_i } \right\|_{\mathbf{R}_i^{ - 1}}^2} 
	+ \kappa \phi(\Delta \mathbf{u}, \Delta \mathbf{x})  \nonumber \\
	\text{s.t.} \;&  \Delta \mathbf{x}_{i+1} = \mathbf{f}_i + \mathbf{A}_i \Delta \mathbf{x}_{i} + \mathbf{B}_i \Delta \mathbf{u}_{i},  i = l, \cdots, k-1 \nonumber
\end{align}
where $\kappa > 0$ is a barrier parameter, and the function $\phi(\cdot)$ is the log barrier defined as  
\begin{align}
\phi(\Delta \mathbf{u}, \Delta \mathbf{x}) 
&= \sum\limits_{i=l}^{k-1} \sum\limits_{j=1}^{n_u} \phi_u (\Delta\mathbf{u}_i(j)) + 
\sum\limits_{i=l}^{k} \sum\limits_{j=1}^{n_x} \phi_x (\Delta\mathbf{x}_i(j)) \nonumber \\
\phi_\star (\Delta\mathbf{\star}_i(j)) & = -\log(\Delta\mathbf{\star}_i^\text{UB}(j) -\Delta\mathbf{\star}_i(j)) \label{eq:phi_star} \\ 
& \quad - \log(\Delta\mathbf{\star}_i(j) - \Delta\mathbf{\star}_i^\text{LB}(j)), \nonumber
\end{align}
with $\star$ representing $\mathbf{u}$ and $\mathbf{x}$, and $j$ referencing to the $j$th entry of the vector $\star$. 
A sequence of the approximate problems (\ref{eq:QP_barrier}) are solved iteratively for a decreasing sequence of values of $\kappa$, as described in Algorithm \ref{alg:primal_barrier}.
For real-time computation, the number of the $\kappa$ values in the sequence is fixed to $n_\kappa$, and we perform $n_{QP}$ iterations for each approximate problem (\ref{eq:QP_barrier}) with a particular value of $\kappa$. 
A simple backtracking line search is used to ensure that the inequality constraints are satisfied at all iterations.

At each iteration in Algorithm \ref{alg:primal_barrier}, 
the Karush-Kuhn-Tucker (KKT) system of the approximate problem (\ref{eq:QP_barrier}) is linearized and solved to compute the search direction represented by $\Delta^2 \mathbf{x}_{i}$ and $\Delta^2 \mathbf{u}_{i}$, ${i=l,\cdots,k}$. 
This is equivalent to solving the following linear MHE problem with only equality constraints, omitting detailed explanations for the sake of brevity:
\begin{equation}\label{eq:kkt_mhe}
	\begin{aligned}
	\mathop {\min }\limits_{\begin{smallmatrix}
		\Delta^2 \mathbf{x}_i, \Delta^2 \mathbf{u}_i
		\end{smallmatrix}} \;
	& \frac{1}{2}\left\| \mathbf{\bar r}_x - {{\Delta^2 \mathbf{x}_l} } \right\|_{\mathbf{P}^{ - 1}}^2
	+ \frac{1}{2}\sum\limits_{i = l}^{k - 1} {\left\| \mathbf{\bar r}_{u,i} - \Delta^2 \mathbf{u}_i \right\|_{\mathbf{\bar Q}_i^{-1}}^2}  \\
	&+ \frac{1}{2}\sum\limits_{i = l}^k {\left\| {\mathbf{\bar r}_{y,i} - \mathbf{\bar C}_i \Delta^2 \mathbf{x}_i } \right\|_{\mathbf{\bar R}_i^{ - 1}}^2} \\
	\text{s.t.} \;&  \Delta^2 \mathbf{x}_{i+1} = - \mathbf{r}_{p,i} + \mathbf{A}_i \Delta^2 \mathbf{x}_{i} + \mathbf{B}_i \Delta^2 \mathbf{u}_{i},  \\
	& i = l, \cdots, k-1. 
	\end{aligned}
\end{equation}
where we define
\begin{subequations}\label{eq:linearize}
\begin{align}
g_{\star, i} &= \text{vect}\left(\left\{ \nabla \phi_\star (\Delta\mathbf{\star}_i^{-}(j)) \right\}_{j=1,2,3}\right), \label{eq:g_star}\\
\mathbf{L}_{\star, i} &= \text{diag}\left(\left\{ \sqrt{\nabla^2 \phi_\star (\Delta\mathbf{\star}_i^{-}(j))} \right\}_{j=1,2,3}\right), \label{eq:l_star} \\
& \qquad\qquad\qquad\qquad\qquad\; \text{with } \star \text{ being } \mathbf{u} \text{ and } \mathbf{x}, \nonumber \\
\mathbf{r}_{p,i} &= - \mathbf{f}_i - \mathbf{A}_i \Delta \mathbf{x}_i^{-} - \mathbf{B}_i \Delta \mathbf{u}_i^- + \Delta \mathbf{x}_{i+1}^-, \\
\mathbf{\bar r}_x &= - \Delta \mathbf{x}_l^-,
\mathbf{\bar r}_{u,i} = \mathbf{Q}^{-1} \left( \mathbf{r}_{u,i} - \Delta \mathbf{u}_{i}^- \right) - \kappa g_{u,i}, \\
\mathbf{\bar r}_{y,i} &= \left[ \begin{matrix}
\mathbf{r}_{y,i} - \mathbf{C}_i \Delta \mathbf{x}_l^- \\
- \sqrt{\kappa} \mathbf{L}_{x,i}^{-1} g_{x,i} 
\end{matrix} \right], 
\mathbf{\bar C}_i = \left[\begin{matrix}
\mathbf{C}_i \\
\sqrt{\kappa} \mathbf{L}_{x,i}
\end{matrix}\right], \label{eq:bar_r_C}\\ 
\mathbf{\bar R}_i &= \text{diag}\left( \mathbf{R}_i, \mathbf{I}_{n_x} \right), 
\mathbf{\bar Q}_i = 
\left( \mathbf{Q}^{-1} + \kappa \mathbf{L}_{u,i}^\text{T} \mathbf{L}_{u,i}  \right)^{-1}.
\end{align}
\end{subequations}
Note that with the scalar entries $\{x(i)\}$,  $\text{vect}(\{x(i)\})$ and $\text{diag}(\{x(i)\})$  in (\ref{eq:g_star}) and (\ref{eq:l_star}) represent a column vector and a diagonal matrix, respectively.
These quantities in (\ref{eq:linearize}) are the result of linearizing the KKT system of the QP (\ref{eq:QP_barrier}).
Inspired by \cite{Have2011}, a structure-exploiting Riccati based approach is adopted to solve the linearized KKT system, which will be detailed in Algorithm \ref{alg:linear_solver}.

\subsection{Implementation aspects}\label{sect:implement}
The implementation using SAO is a critical step for achieving real-time computation of the above algorithm. The following aspects have been considered to either speed up computation or simplify the implementation while maintaining good estimation performance.

The overall computational cost is kept small by setting the horizon length $N$ of the MHE problem (\ref{eq:MHE}) to be 3, the length $N_{\text{eval}}$ of the residual evaluation window in (\ref{eq:rms}) to be 10, the number of iterations $n_\kappa$ and $n_{QP}$ in Algorithm \ref{alg:primal_barrier} to be both two. Extensive numerical simulations show good results even with these small number of iterations.

The AOA or VCAS measurements should not be involved in the proposed MHE algorithm when all redundant AOA or VCAS sensors are identified as faulty. However, this cannot be done by directly removing AOA or VCAS from the output equation (\ref{eq:dt_longoutput}), because the time-varying dimension of the output equation, which implies vectors and matrices of time-varying sizes in the MHE algorithm, cannot be easily handled within the SAO library. To simplify the SAO implementation for the above issue, we let the output equation (\ref{eq:dt_longoutput}) remain the same, but set only the first or third row of the matrix $\mathbf{C}_i$ in (\ref{eq:QPobj}) to be zero after losing all AOA or VCAS sensors, respectively. 
In this way, the feedback information from AOA or VCAS becomes ineffective when necessary, and the SAO implementation still works with vectors and matrices of fixed sizes.

Moreover, some approximations in computing the matrices $\mathbf{C}_i$ and $\mathbf{\bar C}_i$ have been made as explained next.
Let $\mathbf{C}_i(k_1, k_2)$ represent the entry at the $k_1$th row and the $k_2$th column of the matrix $\mathbf{C}_i$.
Over the flight envelop, the two entries $\mathbf{C}_i(2,2)$ and 
$\mathbf{C}_i(3,3)$ are negligible compared to other entries of $\mathbf{C}_i$. Therefore we approximate $\mathbf{C}_i(2,2)$ and 
$\mathbf{C}_i(3,3)$ to be zero. $g_{\star,i}$ and $\mathbf{L}_{\star,i}$ in (\ref{eq:linearize}) and the entries of the third row of $\mathbf{C}_i$ involve logarithm and power computations which are approximated by lookup tables.

The observability of the linearized system in (\ref{eq:QP}) is mainly determined by $\mathbf{C}_i$, because $t_s \nabla_{\mathbf{x}_i} f$ in (\ref{eq:Ai}) is approximately zero and $\mathbf{A}_i$ is very close to an identity matrix for all time instants $i$ over the flight envelope. 
The states become unobservable when all AOA or VCAS sensors are detected to be faulty and removed. This observability issue is due to the fact that all air data sensors are considered to be unreliable and then only inertial sensors can be included in the model described in Section \ref{sect:mdl}. Therefore, the following two scenarios cannot be fully addressed by our method:
\begin{itemize}
	\item When all AOA sensors are removed, neither AOA nor VCAS can be reliably estimated, because all the states $\alpha$, $W_x$, and $W_z$ become unobservable.
	\item When all VCAS sensors are removed, AOA can still be reliably estimated, while VCAS cannot, because $\alpha$ and $W_z$ are still observable, while  $W_x$ becomes unobservable.
\end{itemize}
It should be noted that in the second scenario above, we approximate $\mathbf{C}_i(2,2)$ to be zero as mentioned in the last paragraph, in order to make the MHE algorithm work with the observable subsystem associated with $\alpha$ and $W_z$ and discard the unobservable $W_x$.

\begin{algorithm}[!h]
	\caption{Solve the search direction}
	\label{alg:linear_solver}
	\begin{algorithmic}[1]
		\State \textit{Riccati recursion based factorization}:
		\State $\mathbf{\hat P}_l = \mathbf{P}$
		\For{$i = l \rightarrow k $}
		\State $\Pi_i \leftarrow \mathbf{\bar C}_i \mathbf{\hat P}_i, \; 
		\Xi_i \leftarrow \left(\mathbf{\bar R}_i + \Pi_i \mathbf{\bar C}_i^\text{T}\right)^{-1}, \;
		\Omega_i \leftarrow \mathbf{\bar C}_i^\text{T} \Xi_i$
		\State $\mathbf{K}_i \leftarrow \mathbf{\hat P}_i \Omega_i, \;
		\mathbf{P}_i^f \leftarrow \mathbf{\hat P}_i - \mathbf{K}_i \Pi_i $
		\If{$i<k$}
		\State $\mathbf{\hat P}_{i+1} \leftarrow \mathbf{A}_i \mathbf{P}_{i}^f \mathbf{A}_i^\text{T} + \mathbf{B}_i \mathbf{\bar Q}_i \mathbf{B}_i^\text{T}$
		\EndIf
		\EndFor
		\State \textit{Forward recursion}:
		\State $\Delta^2 \mathbf{\hat x}_l = \mathbf{\bar r}_x$
		\For{$i = l \rightarrow k$}
		\State $\mathbf{\breve{r}}_i \leftarrow {{\mathbf{\bar r}_{y,i}} - {\mathbf{\bar C}_i}\Delta^2 {{\mathbf{\hat x}}_i}}, \; 
		{\Delta ^2}{{\mathbf{x}'}_i} \leftarrow \Delta^2 {{\mathbf{\hat x}}_i} + \mathbf{K}_i \mathbf{\breve{r}}_i$
		\If{$i < k$}
		\State $\Delta^2 {{\mathbf{u}'}_i} \leftarrow \mathbf{\bar r}_{u,i}$
		\State $\Delta^2 {{\mathbf{\hat x}}_{i + 1}} \leftarrow - \mathbf{r}_{p,i} + {\mathbf{A}_i}\Delta^2 {{\mathbf{x}'}_i} + {\mathbf{B}_i}\Delta^2 {{\mathbf{u}'}_i}$
		\EndIf
		\EndFor
		\State \textit{Backward recursion}:
		\State $\Delta^2 \mathbf{x}_k \leftarrow \Delta^2 \mathbf{x'}_k, \;
		\lambda_{k-1} \leftarrow - \Omega_{k} \mathbf{\breve{r}}_{k}$ 
		\For{$i = k-1 \rightarrow 0$}
		\State $\xi_i \leftarrow \mathbf{A}_i^{\rm{T}}{\lambda _i}, \; \Delta^2 {\mathbf{u}_i} \leftarrow \Delta^2 {{\mathbf{u}'}_i} - \mathbf{\bar Q}_i \mathbf{B}_i^{\rm{T}}{\lambda _i}$
		\State $\Delta^2 {\mathbf{x}_i} \leftarrow \Delta^2 {{\mathbf{x}'}_i} - \mathbf{P}_{i}^f \xi_i$
		\If{$i>0$}
		\State ${\lambda _{i - 1}} \leftarrow \xi_i - \Omega_i \left( \mathbf{\breve{r}}_i + \Pi_i \xi_i \right) $
		\EndIf
		\EndFor
	\end{algorithmic}
\end{algorithm}

Computing the search direction by solving the linearized KKT system of (\ref{eq:QP_barrier}) dominates the computational cost of Algorithm \ref{alg:primal_barrier}. This step follows Algorithm \ref{alg:linear_solver} in the SAO implementation by taking the following strategies:
\begin{itemize}
	\item The intermediate results, e.g., $\Pi_i$, $\Xi_i$, $\Omega_i$, $\mathbf{K}_i$ in lines 4-5, $\mathbf{\breve{r}}_i$ in line 13, and $\xi_i$ in line 22 of Algorithm 2, are reused in subsequent computations.
	\item The symmetric or diagonal matrix structures are exploited in all the matrix manipulations.
	\item To compute $\Xi_i$ in line 4 of Algorithm \ref{alg:linear_solver}, the block matrix inversion formula is applied so that the inversion of the $6 \times 6$ matrix $\mathbf{\bar R}_i + \Pi_i \mathbf{\bar C}_i^\text{T}$ can be reduced to the inversion of several $3 \times 3$ matrices and some matrix multiplications. Inverting a $3 \times 3$ matrix is computed via the analytical adjugate formula.
\end{itemize}

\section{Simulation results}
The test scenarios in Table \ref{tab:scenarios} are simulated under both fault-free and faulty conditions using the RECONFIGURE Functional Engineering Simulator \citep{Goup2015, Fern2015}. To mitigate the effect of wind disturbances and model mismatches, the bounds of wind speed and accelerations are set to be 120 kts and 30 kts/s, and the detection thresholds of AOA and VCAS are $J_{\alpha, \text{th}} = 2.3 \text{ deg}$ and $J_{\alpha, \text{th}} = 12 \text{ kts}$, respectively. The simulation results are shown in Table \ref{tab:result}. The estimation performance is evaluated by the absolute estimation errors (AEEs) in their maximum and averaged values, and the detection performance is evaluated by the detection delay for each faulty sensor. In all the fault-free scenarios, the maximum AEEs of AOA and VCAS are less than 0.5 deg and 1.9 kts, respectively, while their mean AEEs are even smaller. There is no false alarm in the fault-free runs.

In the presence of sensor faults, the maximum AEEs become larger compared to those in the corresponding fault-free cases. This is because the faulty sensors are used by the MHE algorithm before they are detected. This effect gets more significant, if the fault amplitude during the detection delay is large, or the detection delay is relatively long, e.g., the nearly 1s detection delay in Scenario 6-F leads to a relatively larger increase in both maximum and mean AEEs. However, the mean AEEs are often less affected by faults than the maximum AEEs, because the estimation errors quickly decrease after isolating the faulty sensors.
Most of the detection delays are less than 0.45s, except for 0.92s in Scenario 6-F. Such short detection delays make sure that the influence of fault information on the estimation performance is small. Therefore, the maximum AEEs of AOA and VCAS are still satisfying, i.e., 0.82 deg and 1.85 kts, respectively, in all the faulty cases.
The only exception is the unacceptable VCAS estimation error in Scenario 2-F. The reason is that all three VCAS sensors are faulty and removed from the MHE algorithm, which makes the horizontal wind unobservable, as explained in Section \ref{sect:implement}.

\begin{figure}
\centering
	\subfigure[Measurements and residuals]{
		\includegraphics[width=0.9\linewidth]{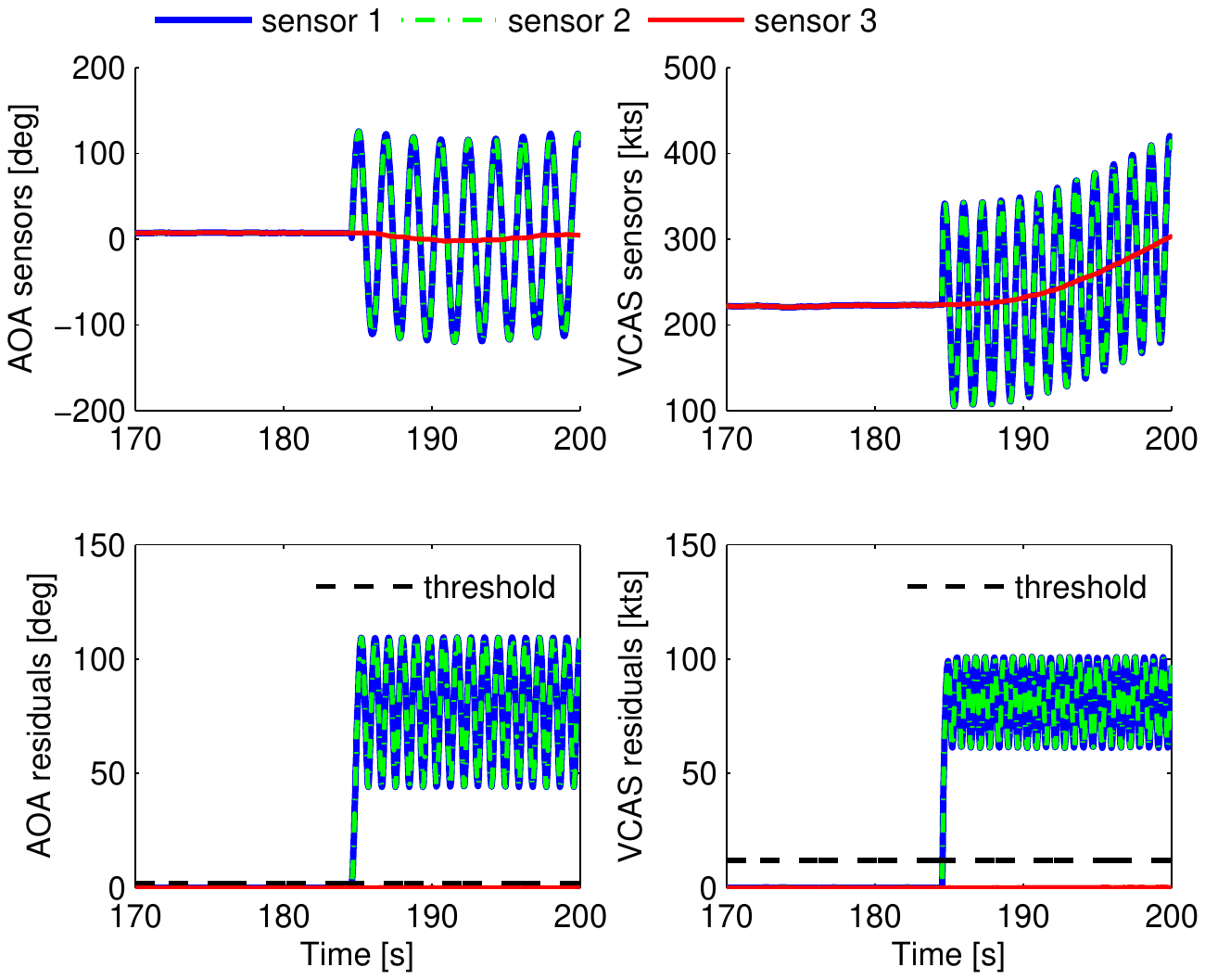}}
	\subfigure[True signals and estimates]{
		\includegraphics[width=0.9\linewidth]{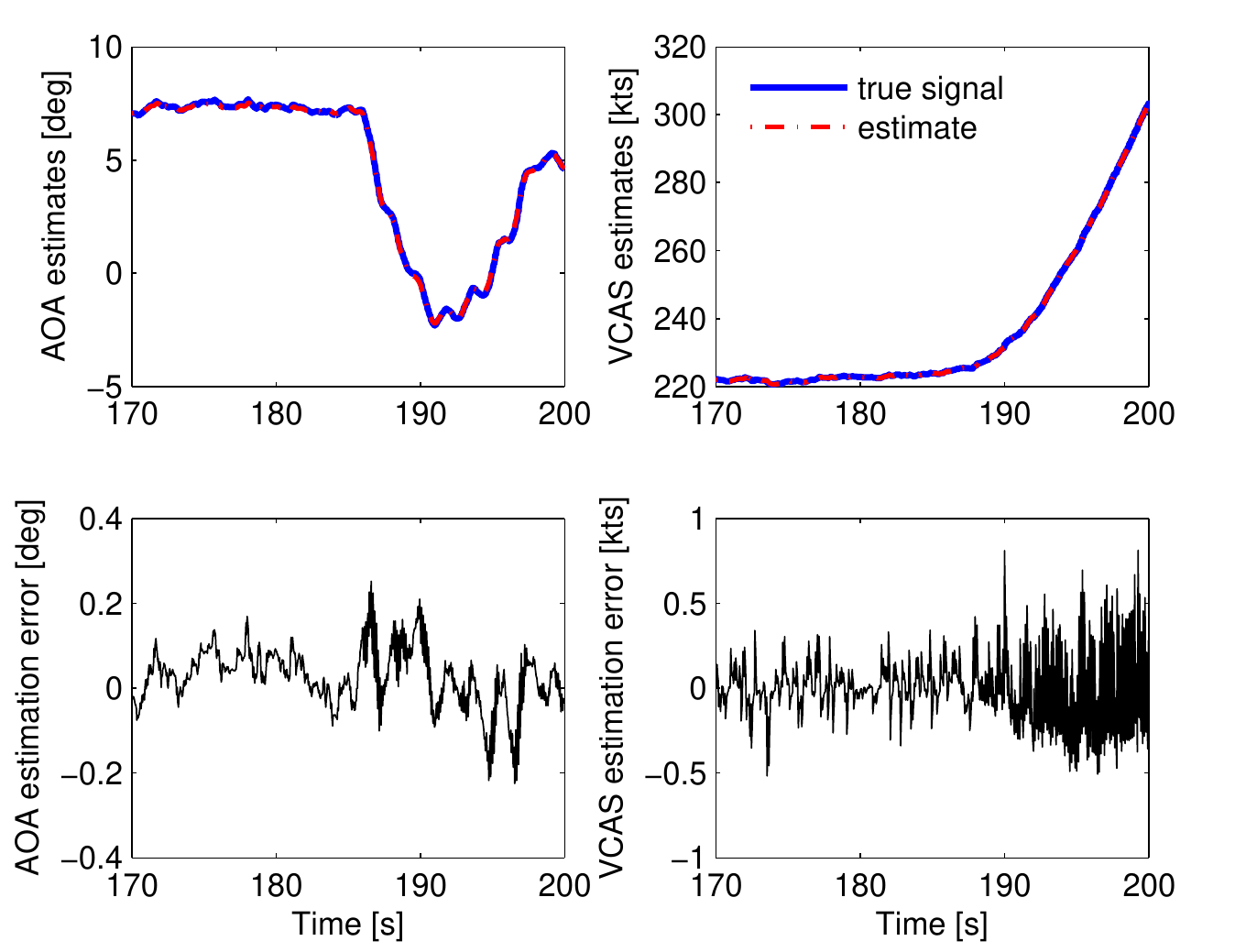}}	
	\caption{Results of Scenario 3-F.}
	\label{fig:result_FPAConLaw_2}
\end{figure}

\begin{figure}
	\centering
	\subfigure[Measurements and residuals]{
		\includegraphics[width=0.9\linewidth]{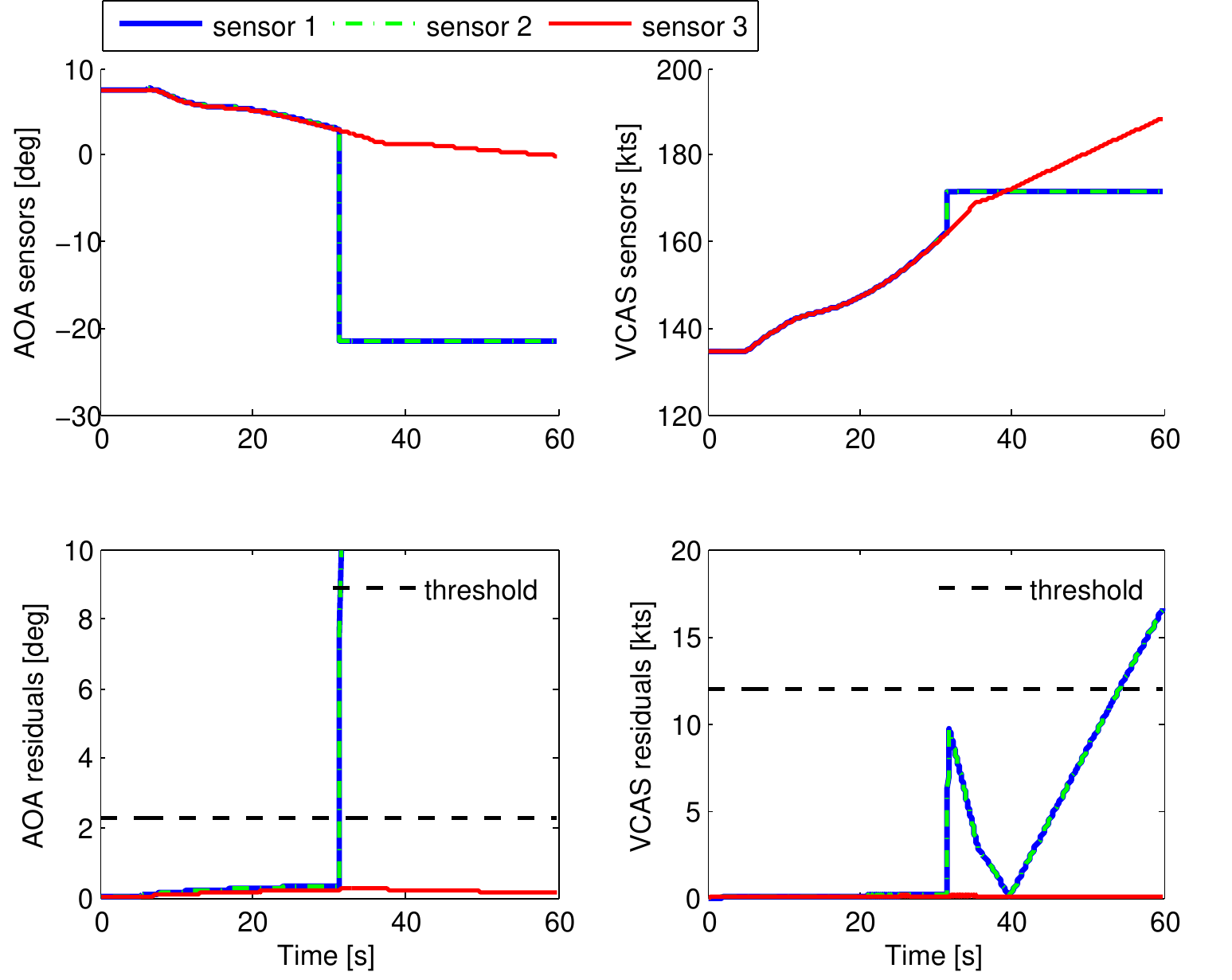}  \label{fig:result_FPAConLaw_2_res}}
	\subfigure[True signals and estimates]{
		\includegraphics[width=0.9\linewidth]{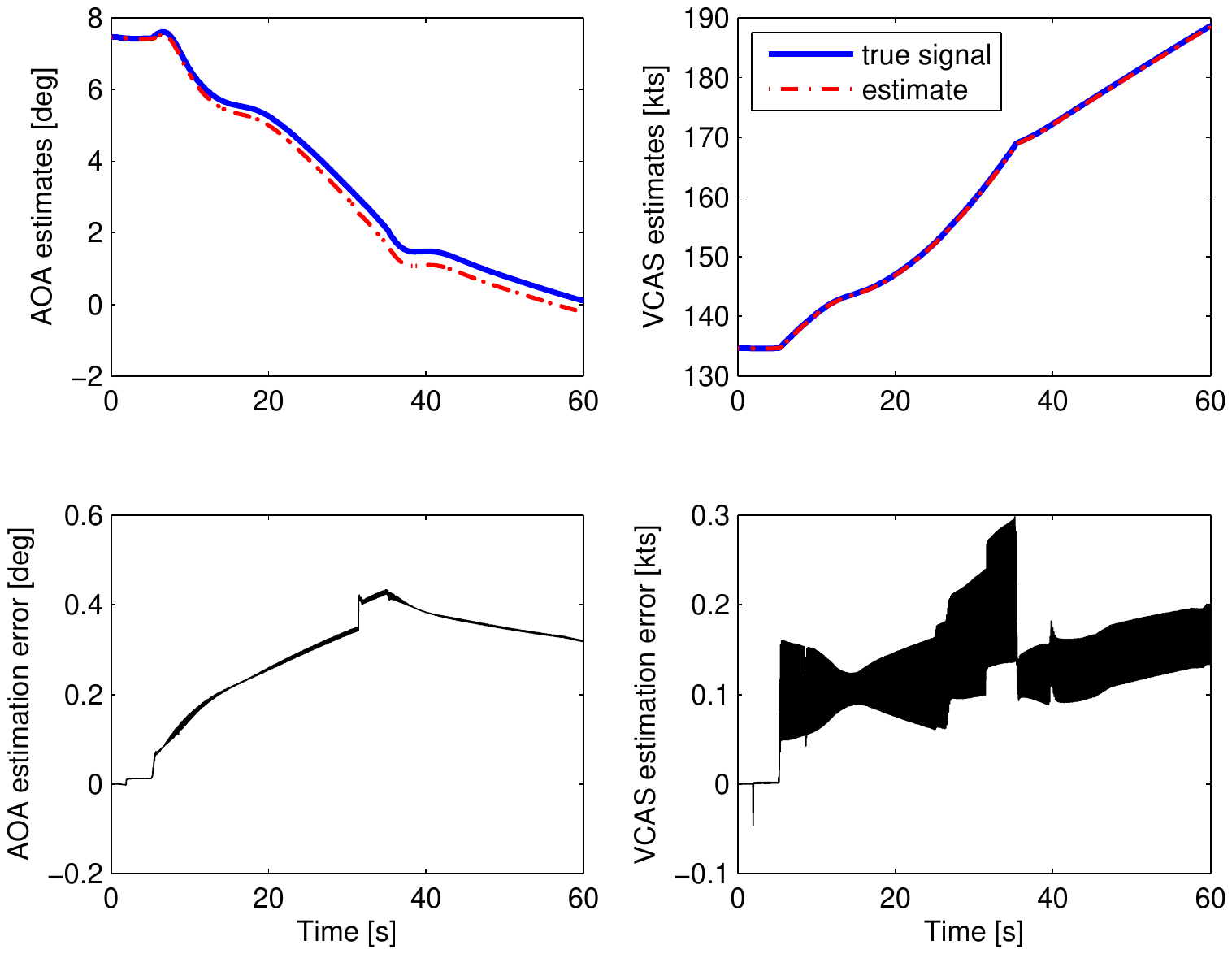}  \label{fig:result_FPAConLaw_2_est}}	
	\caption{Results of Scenario 7-F.}
	\label{fig:result_AOAPR03_21}
\end{figure}

The results of the scenarios 3-F and 7-F are illustrated in Figures \ref{fig:result_FPAConLaw_2} and \ref{fig:result_AOAPR03_21}. In Scenario 3-F, the horizontal and vertical wind disturbances are small, with their peak amplitudes at 1.03 kts. The horizontal wind shear in Scenario 7-F, however, are much stronger, with its peak amplitude at 30.87 kts. It can be seen that our method achieves satisfactory detection and estimation results, even when both AOA and VCAS experience drastic transient changes in the presence of winds and faults.
The relatively long detection delay for the VCAS sensor faults in Scenario 7-F is due to the trade-off between detection delays and false alarms: the detection threshold is determined to avoid false alarms at the cost of increased detection delays. 

The real-time computational cost of our SAO implementation is 5.8 ms per sample under a preliminarily assessment by Airbus. This highlights the feasibility of applying online optimization based MHE methods on FCCs. Future work will focus on further reduction of the algorithm complexity and simplification of the SAO implementation.

\begin{table*}[ht]
	\caption{Simulation results. (In the first column, the indices ``$i$-H'' and ``$i$-F'' correspond to the $i$th maneuver in the fault-free and faulty cases, respectively, as listed in Table \ref{tab:scenarios})}\label{tab:result}
	\centering
	\begin{tabular}{ccccccccccc}
		\toprule
		\multirow{2}{*}{\#} & \multicolumn{4}{c}{Absolute estimation error} & \multicolumn{6}{c}{Detection delay [s]} \\
		& AOA max [deg] & AOA mean [deg] & VCAS max [kts] & VCAS mean [kts] & $\alpha_m^{(1)}$ & $\alpha_m^{(2)}$ & $\alpha_m^{(3)}$ & $V_{m,c}^{(1)}$ & $V_{m,c}^{(2)}$ & $V_{m,c}^{(3)}$ \\		
		\midrule
		1-H & 0.0737 & 0.0500 & 1.7360 & 0.0743 & - & - & - & - & - & - \\
		1-F & 0.1628 & 0.0420 & 1.7748 & 0.1024 & 0.378 & 0.378 & - & 0.378 & 0.378 & - \\
		2-H & 0.1875 & 0.1103 & 1.8532 & 0.1101 & - & - & - & - & - & - \\
		2-F & 0.4938 & 0.1134 & 74.3612 & 42.3752 & 0.368 & 0.368 & - & 0.368 & 0.368 & 0.368 \\
		3-H & 0.4661 & 0.0424 &	0.5368 & 0.0226 & - & - & - & - & - & - \\
		3-F & 0.4661 & 0.0452 &	0.8239 & 0.0349 & 0.433 & 0.433 & - & 0.393 & 0.393 & - \\
		4-H & 0.3643 &	0.0481 & 0.6005 & 0.0300 & - & - & - & - & - & - \\
		4-F & 0.3647 & 	0.0483 & 0.6010	& 0.0300 & - & - & - & 0.363 & 	0.363 & - \\
		5-H & 0.3728 &	0.0524 & 0.6134 & 0.0353 & - & - & - & - & - & - \\
		5-F & 0.3742 &	0.0490 & 0.6111 & 0.0288 & - & - & - & 0.363 & 0.363 & - \\
		6-H & 0.3951 &	0.0473 & 0.6448 & 0.0279 & - & - & - & - & - & - \\
		6-F & 0.6044 &	0.0637 & 1.0238 & 0.0461 & - & - & - & 0.920 & 0.920 & - \\
		7-H & 0.3747 & 0.2530 &	0.2672 & 0.1154 & - & - & - & - & - & - \\
		7-F & 0.4339 & 0.2764 & 0.2978 & 0.1225 & 0.398 & 0.398 & - & 23.278 & 23.278 & - \\
		8-H & 0.3659 & 0.2280 &	0.2547 & 0.1008 & - & - & - & - & - & - \\
		8-F & 0.3629 & 0.2266 &	0.3136 & 0.1167 & - & - & - & 0.373 & 0.373 & - \\
		\bottomrule
	\end{tabular}
\end{table*}

\section{Conclusion}
This paper presents the implementation of our proposed constrained MHE based air data sensor FDI method for industrial V\&V. The involved nonlinear programming at each time instant is solved by the generalized Gauss-Newton sequential quadratic programming strategy and the interior-point method. For real-time computation, we fix the number of iterations per sample. The structure-exploiting Riccati based method is used to solve the linearized KKT system. Specific algorithmic approximations and simplifications are performed to enable its implementation using the Airbus graphical symbol library.
A number of simulation tests on the RECONFIGURE benchmark over different flight points and maneuvers show promising results. Future work will focus on further reduction of the algorithm complexity, fine tuning of algorithm parameters, and extensive Monte Carlo evaluations.


\bibliography{bib_ifac_aca}             

\end{document}